\documentclass[prl,aps,twocolumn,amsfonts,showpacs,superscriptaddress]{revtex4-1}

\newsavebox{\foobox}
\newcommand{\slantbox}[2][0]{\mbox{%
        \sbox{\foobox}{#2}%
        \hskip\wd\foobox
        \pdfsave
        \pdfsetmatrix{1 0 #1 1}%
        \llap{\usebox{\foobox}}%
        \pdfrestore
}}
\newcommand\unslant[2][-.25]{\slantbox[#1]{$#2$}}

\newcommand{\mpi}{\text{\unslant[-.18]\pi}}
\newcommand{\mdelta}{\text{\unslant[-.18]\delta}}

\usepackage{isomath}
\usepackage{graphicx}
\usepackage{xcolor}
\usepackage{rotating}
\usepackage{amsmath,amssymb,graphics}

\usepackage[colorlinks=true, urlcolor=violet, linkcolor=blue, citecolor=red, hyperindex=true, linktocpage=true]{hyperref}

\newcommand\neww[1]{{\color{orange} #1}}

\begin{document}

\title{Absence of disorder-driven metal-insulator transitions in simple holographic models}

\author{Sa\v{s}o Grozdanov}
\email{grozdanov@lorentz.leidenuniv.nl}
\affiliation{Instituut-Lorentz for Theoretical Physics, Leiden University, Niels Bohrweg 2, Leiden 2333 CA, The Netherlands}

\author{Andrew Lucas}
\email{lucas@fas.harvard.edu}

\affiliation{Department of Physics, Harvard University, Cambridge MA 02138, USA}

\author{Subir Sachdev}
\email{sachdev@g.harvard.edu}

\affiliation{Department of Physics, Harvard University, Cambridge MA 02138, USA}
\affiliation{Perimeter Institute for Theoretical Physics, Waterloo, Ontario N2L 2Y5, Canada}

\author{Koenraad Schalm}
\email{kschalm@lorentz.leidenuniv.nl}
\affiliation{Instituut-Lorentz for Theoretical Physics, Leiden University, Niels Bohrweg 2, Leiden 2333 CA, The Netherlands}

\date{\today}

\begin{abstract}
We study electrical transport in a strongly coupled strange metal in two spatial dimensions at finite temperature and charge density, holographically dual to Einstein-Maxwell theory in an asymptotically $\mathrm{AdS}_4$ spacetime, with arbitrary spatial inhomogeneity, up to mild assumptions including emergent isotropy.   In condensed matter, these are candidate models for exotic strange metals without long-lived quasiparticles.    We prove that the electrical conductivity is bounded from below by a universal minimal conductance: the quantum critical conductivity of a clean, charge-neutral plasma.     
Beyond non-perturbatively justifying mean-field approximations to disorder, our work demonstrates the practicality of new hydrodynamic insight into holographic transport.
\end{abstract}

\maketitle

{\it Introduction.---} Novel condensed matter systems such as graphene near half-filling \cite{muller}, cold atomic gases \cite{adams}, and the ever elusive strange metal phase of the high-$T_{\mathrm{c}}$ superconductors \cite{taillefer, keimer, kasahara}, are experimentally realizable strongly interacting many-body quantum systems.
In many of these systems, macroscopic observables are governed by quantum critical physics.  One such observable is the electrical conductance at finite temperature, density and disorder.  Unfortunately, there are few reliable theoretical methods for such strongly coupled regimes.   Some recent work has advocated the memory matrix formalism \cite{zwanzig, mori, forster, lucasMM}, in combination with hydrodynamic insight \cite{hkms}.   These methods may be used directly in microscopic models \cite{raghu, patel, debanjan}.   However, this approach is perturbative in disorder, and may not be adequate for understanding non-Drude physics in strange metals \cite{hartnoll1}.

The advent of gauge-gravity duality \cite{review1, review2, review3}  has allowed for controlled, non-perturbative transport computations in a strongly interacting quantum system.  The quantum dynamics is holographically encoded in the classical response of a black hole in an asymptotically anti-de Sitter space time in one higher dimension.  These systems have large $N$ matrix degrees of freedom, but we focus on correlators of charge currents in this paper.   Correlation functions of simple macroscopic observables are controlled by symmetries and often do not sensitively depend on the underlying large $N$ matrix model.  
  Early efforts to compute transport coefficients entirely from holography were stymied by a simple theorem:
  a Galilean-invariant, charged fluid has zero electrical resistance  (currents are obtained at no energy cost by changing reference frame).  Recently, numerical approaches to general relativity advanced to the point where the transport problem can be solved in numerically constructed black hole backgrounds which break translational symmetry \cite{santoslat1, santoslat2, santoslat3, chesler, ling, santosdisorder1, donos1409, santosdisorder2, rangamani}, giving finite electrical conductivity $\sigma$ at finite density.   Perturbative analytic approaches were developed in \cite{hartnollhofman, btv, lss}; they are equivalent to the memory matrix formalism \cite{lucas2015}.    

To access the ``strongly disordered" regime, massive gravity \cite{vegh, davison, blake1, dsz} and other methods \cite{hartnolldonos, donos1, andrade, gouteraux, donos1406, kim, davison14, davison15, blake15} with similar phenomenology have been developed.  See \cite{karch} for older approaches based on Dirac-Born-Infeld theory.  We refer to these models as ``mean-field disordered", as the geometry is completely homogeneous, despite the fact that momentum relaxes, and so microscopic translational symmetry has been broken.
Holographic mean-field disordered insulators, where $\sigma(T) \sim T^p$ with $p>0$, are not disorder-driven: i) $\sigma$ saturates at a finite value for fixed $T$ independent of the disorder strength (e.g. \cite{donos1406}); ii) insulating behavior in these models is caused by the depletion of charge carriers as $T\rightarrow 0$.   Hence, this broad class of holographic approaches predicts the absence of a disorder-driven metal-insulator transition.  So far, there has been no explicit confirmation of this prediction.

Our letter verifies this  prediction of mean-field disordered models. In doing so, it gives an explicit example of an ``incoherent metal" (but not insulator) in a disordered, isotropic system.   We derive a universal lower bound on the conductivity of certain quantum field theories in two spatial dimensions, which are holographically dual to (classical) Einstein-Maxwell theory, a bulk theory with no free parameters.  Our bound is independent of temperature, the average charge density and its fluctuations.   The new technical developments we exploit are non-perturbative hydrodynamic methods \cite{lucas1506} along with an exact reduction of the holographic transport problem in the Einstein-Maxwell system to a hydrodynamic linear response problem in an inhomogeneous fluid \cite{donos1506}.  Both techniques generalize and we expect that our qualitative result ruling out disorder-driven insulators is generic and holds in more complicated holographic models.

{\it Results.---} We consider classical solutions to the 3+1 dimensional Einstein-Maxwell system with a negative cosmological constant.  The action of this theory is \begin{equation}
S_{\mathrm{bulk}} = \int \mathrm{d}^4x\; \sqrt{-g}\left(\frac{R}{16\mpi G} + \frac{6}{16\mpi GL^2} - \frac{F^2}{4e^2}\right).  \label{eq1}
\end{equation}
$L$ is the AdS radius,  $e$ is the charge of the gauge field,
and $G$ is Newton's constant;  $L=e=
16\mpi G=1$ henceforth.    Note that $e$ is not the unit of charge in the boundary theory \cite{hkms}.   The bulk graviton is dual to the stress-energy tensor of the holographic dual QFT,  and the gauge field is dual to a conserved U(1) (electrical) current.  
We consider a static background geometry \cite{donos1506}:  
\begin{align}
\mathrm{d}s^2 &=  g_{MN}\mathrm{d}x^M\mathrm{d}x^N =
- U(r)V(r,\mathbf{x})\mathrm{d}t^2 \notag \\
&+ U(r)^{-1}W(r,\mathbf{x})\mathrm{d}r^2  + G_{ij}(r,\mathbf{x})\mathrm{d}x_i \mathrm{d}x_j, \label{eq2}
\end{align}
where $\mathbf{x}=(x,y)$ denotes the spatial directions in the boundary theory (denoted with Latin indices $i,j$),  $t$ the time coordinate, and $r$ the bulk radial coordinate.   
 This background is supported by a non-vanishing gauge field: \begin{equation}
A = \Phi(r, \mathbf{x})\mathrm{d}t.  \label{eq3}
\end{equation}
The background is not arbitrary: it obeys the Einstein-Maxwell equations of motion.  
We shall not need its explicit form, but we assume that it asymptotes to AdS at the boundary  $r=\infty$, and that it possesses a connected black hole horizon at $r=0$.  
For such a static black hole, the Hawking temperature $T$ of the black hole must be constant and homogeneous since it has a Killing horizon \cite{wald, visser}. This is the gravitational equivalent of the zeroth law of thermodynamics: an equilibrated system has a constant temperature.
Aside from requiring regularity and constancy of $T$ (all functions are finite at $r=0$, and $U\approx 4\mpi Tr+\ldots$),  bulk fields can have arbitrarily large spatial variations.    
Such spatially variations are sourced by a boundary chemical potential $\phi(\mathbf{x})=\Phi(\infty,\mathbf{x})$, whose explicit form again we shall not need.  We take the spatial dimensions of the dual theory to be compact and flat, with metric $\mdelta_{ij}$: $x$ and $y$ obey periodic boundary conditions $x\sim x+L$ and $y\sim y+L$. We denote (boundary) spatial averages  with $\mathbb{E}[\cdots]$: namely, $\mathbb{E}[f] = L^{-2} \int \mathrm{d}^2\mathbf{x}\; f$.      

The expectation value of the electrical current $\langle j^i\rangle$ in the boundary theory dual to the Einstein-Maxwell system, where $i \in\{x,y\}$, will generically be non-zero when a constant, static, external electric field $E_i$ is applied. In the bulk, we add infinitesimal perturbations to the fields, linear in $E_i$.   The direct current (DC) conductivity $\sigma^{ij}$ is a (boundary) linear response coefficient: \begin{equation}
\mathbb{E}\left[\langle j^i\rangle \right] \equiv J^i= \sigma^{ij} E_j.  \label{eq4}
\end{equation}
As the boundary theory lives on flat space, there is no trouble freely raising and lowering spatial indices  for boundary quantities such as $\sigma^{ij}$.  In the thermodynamic limit $L\rightarrow\infty$, a disordered but on average isotropic theory with time reversal symmetry has $\sigma^{ij}  = \sigma \mdelta^{ij}$.

What we prove in this letter is that for any holographic system with the above assumptions, 
\begin{equation}
\sigma \ge \frac{1}{e^2}=1.  \label{eq6}
\end{equation}
In the boundary theory, the conductivity $\sigma$ is measured in units of $q^2/\hbar$, with $q$ the U(1) charge of $j^i$. Furthermore, $\sigma=1$ exactly  for any uncharged black hole, even with additional bulk matter, so long as it is uncharged under the U(1) sector in (\ref{eq1}), and no higher derivative terms in $F$ are included.  Eq. (\ref{eq6}) is neither valid in more general Einstein-Maxwell-dilaton  theories, where the $F^2$ term in (\ref{eq1}) couples to the dilaton as well, nor in theories with higher derivative terms.   However, if deviations from (\ref{eq1}) are ``small", as they should be in well-defined higher derivative theories, we similarly expect corrections to (\ref{eq6}) to be ``small", and that the conductivities do not vanish when small higher derivative terms are included.

Our results have been anticipated by mean-field treatments of disorder in holography.  A particular example is holographic massive gravity \cite{blake1}, with a consistent graviton mass $m$ added to the bulk action (\ref{eq1}), where \cite{gravfoot} 
\begin{equation}
\sigma  = \frac{1}{e^2}+ \frac{4\mpi \mathcal{Q}_0^2}{\mathcal{S}_0m^2},
\end{equation}
where $\mathcal{Q}_0$ is the homogeneous charge density of the dual theory, and $\mathcal{S}_0$ is the homogeneous entropy density.  This formula also obeys (\ref{eq6}), with the bound saturated on uncharged black holes ($\mathcal{Q}_0=0$).   
This result anticipates (\ref{eq6}), but its reliability for all values of $m$ is surprising.
While the quantitative equivalence between massive gravity and explicitly disordered models has been shown in the limit where bulk deformations which break translation invariance are perturbatively small \cite{btv, lss, lucas2015}, it has been an open question whether or not such models are sensible in the limit of strong disorder.    This letter points out that some predictions of mean-field disorder in  holography are quantitatively correct.   Such justification is useful for a large body of recent holographic work which relies on this approximation to model strong disorder.

A heuristic understanding of our bound follows from hydrodynamic considerations, first emphasized in \cite{lucas1506}:  if the theory thermalizes on short length scales compared to the correlation length of disorder, a theory with locally positive quantum critical conductance cannot be an insulator.    The remarkable results of \cite{donos1506} suggest that this hydrodynamic intuition is mathematically sound in holographic models, regardless of the wavelength of disorder. This formalism, which allows us to derive (\ref{eq6}),  may be a consequence of the holographic large $N$ limit which neglects quantum fluctuations, and has no lattice UV cutoff.  Our result contrasts  with conventional lattice realizations of quantum critical models, which are always insulators at strong enough disorder.  In lattice models, our bound may be realized at intermediate disorder scales, and our theory may be compared in this regime with quantum Monte Carlo computations of the conductivity $\sigma$ \cite{wallin}.

Our holographic models may serve as a candidate theory for an incoherent metal with badly-broken translational symmetry \cite{hartnoll1}.   Incoherent metals are proposed states whose conductivities are intrinsically bounded from below.   Remarkably, our holographic models do have bounded electrical conductivity.    \cite{hartnoll1} argued that the disappearance of a Drude peak in the finite frequency conductivity of some experimental systems may be evidence for such an incoherent metal.    Our work demonstrates that Einstein-Maxwell systems transition from coherent metals (where momentum is long-lived and there is a sharp Drude peak in the conductivity) to incoherent metals,  which remain conductors despite substantial local inhomogeneities.   This defies the conventional phenomenology for strong disorder, and indicates the potential utility of our approach.  To make further connection to experimentally realized bad metals, $\sigma \sim T^{-1}$ is required \cite{kasahara}, instead of $\sigma \sim T^0$.    Our explicit realization of incoherent metals serves as an important first step.

Finally, we note that finely tuned models with disconnected black hole topologies \cite{anninos, santos} exist, but we do not allow for this: randomly disordered black holes appear to stay connected and have uniform temperature \cite{santosdisorder1, santosdisorder2}.   A fragmented black hole may be a holographic many-body localized \cite{basko} state, and may evade our bound.

{\it 
  Technical Details.---}  The remainder of this letter contains the proof of (\ref{eq6}).  We begin by re-interpreting physically the results of \cite{donos1506}.   They found, following the general membrane paradigm technique of \cite{iqbal, donos1409},  that if the asymptotic expansion of $G_{ij}$ and $\Phi$ near the horizon is \begin{equation}
\Phi = r S(\mathbf{x})\mathcal{Q}(\mathbf{x})
+ \cdots, \;\;\; G_{ij} = \gamma_{ij}(\mathbf{x}) + \cdots, 
\end{equation} with $S(\mathbf{x}) = V(0,\mathbf{x})$, whilst a constant infinitesimal electric field $E_j$ (note index is lowered) and its thermal analogue $\zeta_j$ (similar to to $-\partial_j \log T$) are applied to the boundary theory, then the following equations hold (we change notation from \cite{donos1506}
): 
\begin{subequations}\label{eq9}\begin{align}
&\!\! \nabla_i \left(T \mathcal{S}  v^i\right) = \nabla_i \left(\mathcal{Q} v^i + \sigma_{\textsc{q}}\left(E^i - \partial^i  \mu\right)\right) = 0,  \label{eq9a} \\
&\!\! \mathcal{Q}(E_j - \partial_j  \mu) + \mathcal{S}(T\zeta_j-\partial_j  \Theta)  +  2\eta \nabla^i \nabla_{(i}  v_{j)} =0 , \label{eq9b} 
\end{align}\end{subequations}
with covariant derivatives taken with the metric $\gamma_{ij}$.   The coefficients $ \mu$, $ \Theta$ and $ v_i$ are interpreted as the chemical potential, temperature (difference) and velocity of an emergent horizon fluid, and may be expressed in terms of perturbations of the bulk fields \cite{donos1506}; their specific forms are irrelevant for our purposes.
$\eta = \sigma_{\textsc{q}}= 1$ and $\mathcal{S}=4\mpi$ are constants.   It is, however, helpful to not substitute these explicit values in just yet, as in the form (\ref{eq9}) it is straightforward how to apply the formalism of \cite{lucas1506}.    Eqs. (\ref{eq9a}) encode heat and charge conservation in a fluid with a constant entropy density $\mathcal{S}$, variable charge density $\mathcal{Q}$, and constant isotropic quantum critical thermoelectric conductivities $\sigma_{\textsc{q}}=1$, $\alpha_{\textsc{q}}=\bar\kappa_{\textsc{q}}=0$, living on a curved space with metric $\gamma_{ij}$.   Eq. (\ref{eq9b}) is the analogue of the Navier-Stokes equation for an isotropic, relativistic fluid with shear viscosity $\eta$ (the value of bulk viscosity is not important, due to incompressibility (\ref{eq9a})).  

Consider now the spatial average of the currents associated with this horizon fluid:
\begin{subequations}\begin{align}
J^i  &= \mathbb{E}\left[\sqrt{\gamma}\mathcal{Q}v^i + \sigma_{\textsc{q}}\sqrt{\gamma}\gamma^{ij}\left(E_j - \partial_j  \mu\right)\right], \label{jhor} \\
 Q^i &= \mathbb{E}\left[\sqrt{\gamma}T\mathcal{S} v^i\right].
\end{align}\end{subequations} 
The Einstein-Maxwell equations imply that these spatial averages are independent of the holographic radial direction and that they encode the   spatially averaged charge and heat currents of the boundary theory.
We can therefore compute transport coefficients in the disordered boundary theory by solving a linear response problem in the disordered horizon fluid for the electrical current $J^i$ and heat current $Q^i$ as a function of $E_i$ and $\zeta_j$.  

The holographic DC transport problem has thus reduced to a hydrodynamic computation in the framework of \cite{lucas1506} (after a straightforward generalization to curved space).   One direct consequence of hydrodynamics is Onsager reciprocity:  the  
thermoelectric conductivity matrix is symmetric.  This is distinct from positive-definiteness
, which was shown in \cite{donos1506}.

We first compute the conductivity of uncharged black holes, where $\mathcal{Q}=0$. 
Generically, this will only occur when $A=0$ in the background solution.     In this case charge and heat transport decouple and (\ref{eq9}) simplifies to \begin{equation}
\partial_i \left(\sqrt{\gamma}\gamma^{ij}\left(E_j - \partial_j  \mu\right)\right) = 0, \label{eq10}
\end{equation}
with $\Theta=v^i=0$ consistently set to vanish.   (\ref{eq10}) has a unique solution for $\mu$, up to an unimportant constant, and is also unchanged by additional bulk matter so long as it does not couple to the Maxwell term in (\ref{eq1}).

As (\ref{eq10}) is linear, we may write $\mu = \mu^j E_j$.  It follows that for some constants $\psi_i^j$ and a (periodic) function $\Psi^i$: \begin{equation}
\sqrt{\gamma}\gamma^{ij}\left(\mdelta^k_j - \partial_j  \mu^k\right) = \epsilon^{ij}\left(\psi_j^k  - \partial_j \Psi^k\right)  \label{eq12}
\end{equation}with $\epsilon^{xy} = -\epsilon^{yx}=1$, $\epsilon^{xx}=\epsilon^{yy}=0$.  (\ref{eq12}) is equivalent to 
\begin{equation}
-\epsilon^{ij}\left(\mdelta_j^k - \partial_j \mu^k\right) = \sqrt{\gamma}\gamma^{ij}\left(\psi^k_j - \partial_j\Psi^k\right), \label{eq14}
\end{equation}and taking another spatial derivative: \begin{equation}
\partial_i \left(\sqrt{\gamma}\gamma^{ij}\left(\psi^k_j - \partial_j\Psi^k\right)\right)=0.
\end{equation}Uniqueness and linearity fix \begin{equation}
\partial_i\Psi^k = \psi^k_j \partial_i \mu^j . \label{eq16}
\end{equation}Combining (\ref{eq14}) and (\ref{eq16})\neww{,} we find \begin{equation}
\sqrt{\gamma}\gamma^{ij} \left(\mdelta^k_j - \partial_j \mu^k\right) = - \epsilon^{ij} \left(\mdelta^m_j - \partial_j \mu^m\right) \left(\psi^{-1}\right)^k_m. \label{eq17}
\end{equation}
Contracting with $E_k$ and integrating over the torus, the left hand side equals $J^i$.  Using (\ref{eq4})\neww{,} we find \begin{equation}
\epsilon^{ik}\psi^j_k = \sigma^{ij}.
\end{equation} 
Eqs. (\ref{eq12}) and (\ref{eq17}) are only consistent if \begin{equation}
\det(\sigma^{ij})=1.  \label{deteq}
\end{equation}
This holds for any theory, regardless of the assumption of isotropy.   But if we further assume isotropy, we conclude  \begin{equation}
\sigma=1.  \label{sigma1}
\end{equation}
This confirms that the lower bound of (\ref{eq6}) is satisfied whenever the black hole is uncharged.   

Eq. (\ref{sigma1})  is reminiscent of a known result that random resistors arranged on a square lattice, with the logarithmic resistances symmetrically distributed around 0, have an exact conductivity of 1 in the thermodynamic limit \cite{halperin}.     If we suppose for simplicity that $\gamma_{xy}=0$, then a discretized approximation to the diffusion equation (\ref{eq10}) would consist of resistors on a square lattice of resistance $R=\sqrt{\gamma_{yy}/\gamma_{xx}}$ when oriented in the $x$-direction, and resistors of resistance $R=\sqrt{\gamma_{xx}/\gamma_{yy}}$ when oriented in the $y$-direction.   In a theory which is on average isotropic, it is clear that $\log R$ is symmetrically distributed around 0, for every resistor.    This gives a heuristic justification for (\ref{sigma1}).   Such exact results for resistor networks are rare.   Likewise, we do not expect the generalization of (\ref{sigma1}) to higher dimensional uncharged black holes to exactly match the non-perturbative prediction of massive gravity.

Bulk Maxwell self-duality \cite{son} provides further physical insight into  (\ref{deteq}).   Define a bulk dual Maxwell tensor \begin{equation}
\mathcal{F}^{MN}  =- \frac{1}{2}\varepsilon^{MNPQ}F_{PQ} ,
\end{equation}with $\varepsilon^{rtxy} = 1/\sqrt{-g}$.  Note that: 
\begin{equation}\label{Fduality}
\mathcal{F}^{rj} = \frac{1}{\sqrt{-g}} \epsilon^{ji} F_{ti} 
\end{equation}
and that as $r\rightarrow\infty$, \eqref{Fduality} implies that (from the dual Maxwell tensor) there is a constant dual current density $\mathcal{I}^i=\epsilon^{ij}E_j$ in the boundary theory.  $\nabla_M F^{Mi}=0$, whose spatial average leads to $\partial_r J^i=0$, is the Bianchi identity for $\mathcal{F}$, and implies $\partial_r\mathbb{E}[\mathcal{F}_{it}]=0$.  Using $\mathcal{F}$ at $r=0$, \cite{indexnote}
\begin{equation}
\mathbb{E}[\mathcal{F}_{it}] \equiv \mathcal{E}^i = \epsilon^{ij} J^j.
\end{equation}
The spatially averaged dual electric field $\mathcal{E}_i$ is independent of bulk radius and may be evaluated at the horizon. Denoting with $\mathfrak{r}$ the dual resistivity tensor, we conclude that $\mathcal{E}^i = \mathfrak{r}^{ij}\mathcal{I}^j$.
Relating $\mathcal{E}$ and $\mathcal{I}$  to $E$ and $J$ leads to 
\begin{equation}
J^i = -\epsilon^{ij}\mathfrak{r}^{jk}\epsilon^{kl}E^l.  \label{frakeq2}
\end{equation}
If we assume that the disordered boundary theory is particle-vortex dual, then we expect $\mathfrak{r}^{-1}=\sigma$.   This assumption, along with (\ref{frakeq2}), leads to (\ref{deteq}).    Theories with holographic duals with bulk Maxwell electromagnetism thus appear to have remnants of particle-vortex duality even when translational symmetry is strongly broken.  

We now prove the bound (\ref{eq6}) for charged black holes using a variational technique  from \cite{lucas1506}.   In our holographic model, the following inequalities are satisfied: 
\begin{align}
  \frac{\bar{\mathcal{J}}^2 }{\sigma}  &\le \frac{\bar{\mathcal{J}}^2  T\bar\kappa }{T\sigma\bar\kappa -T^2\alpha^2} 
\le \mathbb{E}\left[ 2 \nabla^{(i}\mathcal{V}^{j)} \nabla_{(i}\mathcal{V}_{j)} \sqrt{\gamma} \right. \notag \\
&\;\;\;\;\;\; \left.+  \left(\mathcal{J}^i - \mathcal{Q}\mathcal{V}^i\right)\left(\mathcal{J}_i - \mathcal{Q}\mathcal{V}_i\right)\sqrt{\gamma}\right],   \label{eqb}
\end{align}
with $\bar\kappa$ the thermal conductivity when $E_i=0$, $\alpha$ the Seebeck coefficient, both assumed to be isotropic, and $\mathcal{J}^i$ and $\mathcal{V}^i$ arbitrary vector fields such that 
\begin{align}
\nabla_i \mathcal{V}^i &= \nabla_i \mathcal{J}^i = 0,  \label{eqs} \\
\bar{\mathcal{J}}^i &= \mathbb{E}\left[\sqrt{\gamma} \mathcal{J}^i\right],\qquad \mathbb{E}\left[\sqrt{\gamma} \mathcal{V}^i\right] = 0.  \label{eq22}
\end{align}
Note that $\bar{\mathcal{J}}^i$ is the averaged current measured in the boundary theory on our trial current $\mathcal{J}^i$, and so the index in $\bar{\mathcal{J}}^i$ is raised and lowered with $\mdelta_{ij}$.  
The second inequality in (\ref{eqb}) is satisfied only by the exact solutions of the equations of motion, when 
\begin{equation}
\mathcal{V}^i = v^i, \qquad \mathcal{J}^i = \mathcal{Q}v^i + E^i - \partial^i \mu.
\end{equation}   
Furthermore, the last constraint in (\ref{eq22}) ensures there is no net heat flow \cite{lucas1506}.     

The inequality (\ref{eqb}) follows from studying the power that would be dissipated if we tried to force electrical current $\mathcal{J}^i$, and velocity $\mathcal{V}^i$, through the fluid.  The fact that the dissipated power is minimized by the true solution to the equations of motion is analogous to the fact that the power dissipated in a resistor network, allowing arbitrary current flows through each resistor, up to current conservation, is minimzed on the true solution where both of Kirchoff's Laws are obeyed \cite{levin}.

For any charged black hole, we may set $\mathcal{V}^i=0$, which trivially satisfies our constraints.   For $\mathcal{J}^i$, we choose 
\begin{equation}
{\mathcal{J}}^i = E^i - \partial^i  \tilde\mu \equiv \tilde{\mathcal{J}}^i ,  \label{eq33}
\end{equation}
where $\tilde\mu$ is the exact solution to the diffusion equation (\ref{eq10}), using the metric $\gamma_{ij}$ of the charged black hole.    $\tilde{\mathcal{J}}^i$ is the exact electric current that would flow upon application of the electric field, if we remove all charge density from the horizon fluid, but keep the same $\gamma_{ij}$.   
If $\tilde{\mathcal{J}}^i$ is normalized by (\ref{eq22}), \begin{equation}
\bar{\mathcal{J}}^2 = \mathbb{E}\left[\sqrt{\gamma} \tilde{\mathcal{J}}^i\tilde{\mathcal{J}}_{i}\right].  \label{eqlast}
\end{equation}
To derive this result, we use that $\alpha=0$ for an uncharged black hole, along with (\ref{sigma1}) and the fact that the bound (\ref{eqb}) is saturated on the true electrical current for the uncharged black hole, which is $\tilde{\mathcal{J}}^i$, along with $\mathcal{V}^i=0$.   The ansatz $\mathcal{J}^i=\tilde{\mathcal{J}}^i$ and $\mathcal{V}^i=0$ is no longer the true solution when $\mathcal{Q}\ne 0$, but we may still use it as a variational ansatz;  however, in this case, we expect $\alpha\ne 0$, and that each inequality in (\ref{eqb}) is strict.  
Using (\ref{eqlast}) in (\ref{eqb}), we obtain our main result, (\ref{eq6}).

\textit{Acknowledgements.---} We thank R. Davison, B. Gout\'eraux, B. Halperin, S. Hartnoll and J. Santos for helpful discussions. 
AL and SS are supported by the NSF under Grant DMR-1360789, the Templeton foundation, and MURI grant W911NF-14-1-0003 from ARO.
Research at Perimeter Institute is supported by the Government of Canada through Industry Canada 
and by the Province of Ontario through the Ministry of Research and Innovation.   SG and KS are supported by a VICI grant of the Netherlands Organization for Scientific Research (NWO), by the Netherlands Organization for Scientific Research/Ministry of Science and Education (NWO/OCW) and by the Foundation for Research into Fundamental Matter (FOM).


\begin{thebibliography}{99}
\bibitem{gravfoot}  Similar results are found using other mean field ways to model disorder in Einstein-Maxwell theory \cite{donos1, andrade}.
\bibitem{indexnote} Recall that for boundary theory quantities such as those referenced below, indices may be raised and lowered freely with $\mdelta_i^j$.
\end{thebibliography}
\end{document}